# A new FDSOI spin qubit platform with 40nm effective control pitch

T. Bédécarrats[1], B. Cardoso Paz[2], B. Martinez Diaz[3], H. Niebojewski[1], B. Bertrand[1], N. Rambal[1], C. Comboroure[1], A. Sarrazin[1], F. Boulard[1], E. Guyez[1], J.-M. Hartmann[1], Y. Morand[1], A. Magalhaes-Lucas[1], E. Nowak[1], E. Catapano[1], M. Cassé[1], M. Urdampilleta[2], Y.-M. Niquet[3], F. Gaillard[1], S. De Franceschi[3], T. Meunier[2], M. Vinet[1]

Université Grenoble Alpes and [1] CEA-Leti, [2] CNRS Institut Néel, [3] CEA-Irig, F-38000 Grenoble, France
Email: thomas.bedecarrats@cea.fr  heimanu.niebojewski@cea.fr

*Abstract—* Operating Si quantum dot (QD) arrays requires homogeneous and ultra-dense structures with aggressive gate pitch. Such a density is necessary to separately control the QDs chemical potential (i.e. charge occupation of each QD) from the exchange interaction (i.e. tunnel barriers between each QD). We present here a novel Si quantum device integration that halves the effective gate pitch and provides full controllability in 1D FDSOI QD arrays. The major advantages of this architecture are explored through numerical simulations. Functionality of the fabricated structure is validated via 300K statistical electrical characterization, while tunnel-coupling control is demonstrated at cryogenic temperature.

## I. Introduction

Silicon spin qubits are very promising candidates in the quest for quantum computing [1,2]. One of the major advantage Si offers in comparison to other platforms is its scalability. Making millions of qubits, similarly to making millions of transistors, should enable the use of quantum error correction algorithms, resulting in error-free quantum computing [3]. However, a high density of qubit devices infers that at least as many ways of control over those qubits are required, which is challenging from an integration perspective.

FDSOI qubit devices with gate pitches as low as 64nm have been demonstrated in previous work using hybrid deep UV(DUV)/e-beam lithography [4]. Such a device is shown in Fig. 1 left. It includes a top gate electrode at the metal 1 (M1) level that provides control over the tunnel coupling between two consecutive QDs, in addition to the backgate electrode below the BOX enabled by the SOI substrate. However, only a very weak modulation of the device drain current is measured when sweeping the top gate polarization over a wide voltage range (Fig. 1 right). This is due to the too large distance between M1 metal lines and the qubit active layer, combined with electrostatic screening from the front gates even when the gate pitch is relaxed. Implementing global exchange gates far from the QDs is therefore not an effective solution to control exchange interactions for large QD arrays.

In this paper, we investigate the integration of local exchange gates (J-gates). In comparison to a global top gate (Fig. 2), our J-gates design consists in metallic trenches that intertwine with front-gates to achieve independent tunability between QD charge occupation and tunnel barriers (Fig. 3). The resulting structure has an effective controllability pitch that is therefore half that of the front-gates, while fabrication flow deviates very little from standard CMOS technology.

In the first part of this work, we investigate the combined effect of J-gates and backgate electrodes through numerical simulations. We then detail the process steps yielding to a 4-gates linear QD array, followed by electrical validation at room temperature. Finally, we present electrical data at cryogenic temperature demonstrating the exchange interaction.

## II. Simulation

A proof-of-concept was given by Poisson + effective mass calculations on linear arrays of $2\times N$ face-to-face (F2F) dots (Fig. 4). Dots are controlled by a first level of front gates that partly overlap the channel, while the tunnel couplings between the dots are controlled by a second (intertwined) level of exchange gates that run over the whole width of the channel. In such an array, neighboring dots on one side of the channel are qubits, while the dots on the other side are used for readout purposes. Therefore, the design must be versatile enough to switch from a configuration where dots are coupled longitudinally for two-qubit operations, to a configuration where they are F2F coupled for readout. This can actually be achieved with a combination of exchange and back gates settings, as shown in Fig. 5. On the one hand, an increasingly positive exchange gate bias $V_J$ enhances both the longitudinal tunnel coupling $t_{//}$ between neighboring dots and the transverse tunnel coupling $t_\perp$ between F2F dots. On the other hand, an increasingly negative $V_{bg}$ squeezes dots in the corners of the channel, suppressing the tunneling $t_\perp$ between F2F dots (Fig. 5), and yielding two-qubit ($V_{bg} \ll 0$ V) as well as readout ($V_{bg} \approx 0$ V) functionalities.

One of the main challenges in the design and fabrication of exchange gates is to achieve a tight enough electrostatic control. Exchange gates indeed tend to be screened by the front gates underneath. The efficiency of the exchange gates ($\partial t_{//}/\partial V_J$ or $\partial t_\perp/\partial V_J$) can be improved by thinning the front gates (Fig. 6) or relaxing the pitch (larger inter-dot spacing).

## III. Technology

Quantum devices were processed on Leti 300mm FDSOI line using only immersion DUV lithography. In addition to being fully compatible with industry standards [4], this solution provides better throughput and thus faster learning over other integration schemes relying on hybrid DUV/e-beam lithography [1,6].

Our Si quantum integration is based on a standard FDSOI CMOS flow. We implemented minor adjustments to our standard FDSOI transistor process flow and developed a new "QTrench" module for the fabrication of exchange gates, as shown in Fig. 7.

Starting from SOI wafers, silicon nanowires are patterned in a MESA isolation scheme. Gate patterning is realized by single patterning exposure using immersion 193nm DUV lithography. A gate pitch as low as 80nm is achieved, resulting in very dense and homogeneous linear gate arrays (Fig. 8). In order to maximize the J-gate efficiency –as suggested by simulation results from Fig. 6, this patterning strategy was validated on an optimized TiN/poly-Si gate stack, where the overall gate height was reduced by 18nm (Fig. 9). Next, dielectric spacers are formed in the inter-gates regions to prevent dopants from penetrating into the underlying silicon layer during the upcoming fabrication of reservoirs. The latter is obtained by selective epitaxial growth of Si:P or $Si_{0.7}Ge_{0.3}$:B heavily in-situ doped layers, depending on whether electrons or holes qubits are considered. Following NiPt silicidation and dielectric encapsulation, contacts on gates and reservoirs are plasma etched. J-gates are then formed by etching trenches that intertwine with the previous gate level. The device shown in Fig. 10 has 80nm pitch front-gates and 80nm pitch J- gates, resulting in an effective 40nm gate pitch controllability. Thanks to the optimized front-gate height, a distance as low as 20nm is obtained between exchange gates and the silicon qubits layer. This is achieved with no penalty on leakage, as demonstrated in the next section. Finally, contact metallization and BEOL modules are processed for terminals routing.

## IV. ROOM TEMPERATURE MEASUREMENTS

In order to validate electrically our exchange gate module, we first performed statistical measurements of the J-gate leakage currents on more than 2500 devices. Devices with 2-QDs in series (i.e. 2 front gates + 3 J-gates in series above a silicon nanowire) exhibit a 98.3% isolation yield (Fig. 11).

Extensive characterization was also performed on larger 1D arrays, with 4 QDs/5 J-gates in series, similar to the device shown in Fig. 10. We selected 90nm pitch devices to have enhanced J-gate control. Among the 384 measured devices, we selected 79 devices showing full functionality (see Fig. 12). The typical transistor behavior of their front gates is extracted by successive measurements of the drain current versus each front gate voltage. Threshold voltages $V_T$ are, on average, almost identical between each gate in the array, with a mean value around 0.41V, which is consistent with the metal-gate stack (2.5nm SiO2/TiN) used (Fig. 13). At the single device scale, the matching between all possible gate pairs in a 4 QDs device is shown in Fig. 14. The low variability of the ΔVT distribution demonstrates the high homogeneity of the gate patterning. Matching parameters AVT are even lower than that of 1-gate-level hybrid DUV/e-beam devices [7]. A slight discrepancy is evidenced in the subthreshold slope (SS) values between the various gates of the device (Fig. 15). The larger outer gate SS range is attributed to the intrinsic variability of the source drain junctions present in their neighborhoods.

Let us now focus on the impact of J-gates. To this end, the front gates are set to 0.5V, slightly above their threshold voltage, in order to maximize the impact of the J-gate polarization on the drain current. $I_D(V_J)$ characteristics on a [-10V +10V] range are shown in Fig. 16, demonstrating a clear $I_D$ modulation over 6 decades. This behavior is well reproduced by TCAD simulations. The calibrated deck that was developed enables to (i) easily benchmark different layouts for front and J-gates and (ii) quantify the impact of the gate pitch (see Fig. 17). In particular, the dependence of the SS on the gate pitch from simulations is in very good agreement with experimental data (Fig. 18).

## V. LOW-T MEASUREMENTS

2 QDs/3 J-gates in series devices were wire-bonded on a printed circuit board (PCB) and mounted on a home-made dipstick operating at 4.2K. Fig. 19 (a) shows the bias configuration used in order to obtain a double dot regime, with one QD below each gate. Access gates (J1 and J3) are biased at -5V to increase barriers between QDs and source/drain reservoirs, while a negative voltage at J2 can be adjusted to modulate the coupling between QD1 and QD2. A back gate bias of +25V was applied to push QDs close to the back interface (Si/BOX), where the number of defects are expected to be smaller and noise to be reduced, because of an improved interface quality w.r.t. the front interface [8,9]. $I_{DS}$-$V_{GS}$ curves (Fig. 19.b) at low drain bias (0.5mV) show characteristic Coulomb peaks, whose position varies according to the applied $V_{J2}$. For QD1, the charging energy ($E_C$), the lever arm (α) and elements of the capacitance matrix that describes a single electron transistor [10] were extracted from measurements of Coulomb blockade diamonds (Fig. 20). Five diamonds can be distinguished. The first one has $E_C$ = 7.6meV and α = 0.34. As the QD gets bigger (i.e. the number of electrons increases), $E_C$ and α become lower, while the gate capacitance remains about 8 aF for all diamonds.

Finally, a sequence of stability diagrams measured at different $V_{J2}$ (Fig. 21) confirms the effectiveness of the exchange gates and their capability of tuning the coupling between QD1 and QD2. A single-dot turns into a double-dot system when sweeping $V_{J2}$ from 0V towards more negative values (see the characteristic honeycomb pattern when $V_{J2}$ = -6V). The same voltage range was used for G1 and G2 in Fig. 21, indicating that the two QDs must be rather similar in shape/size, as expected from the good matching data obtained at room temperature.

## VI. CONCLUSION

We successfully demonstrated a new FDSOI spin qubits platform with industry-compatible process including two gate layers, resulting in an effective control pitch down to 40nm. Room temperature measurements showed the low variability of the electrical characteristics of the devices fabricated with this platform, which is promising for scaling to larger systems. The electrostatic control over 2-gate devices was successfully achieved at 4.2K. As predicted by numerical simulations, the use of exchange gates together with a positive back biasing enabled to control the coupling between 2 QDs formed below each gate. This is the first demonstration of electrostatic coupling control over QD systems implemented in CMOS SOI devices by means of back biasing and the use of exchange gates. It is a first step towards a successful control of spins for qubit applications.

ACKNOWLEDGMENT

This work was partly supported by the EU through the H2020 QLSI project and the European Research Council (ERC) Synergy QuCube project.

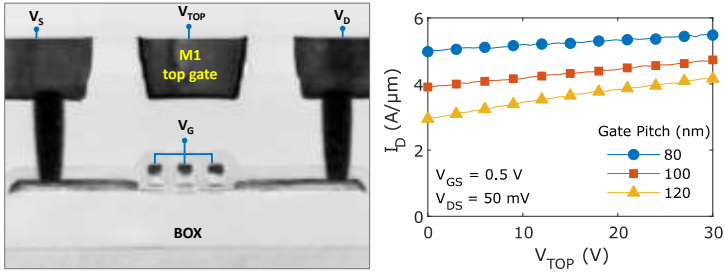
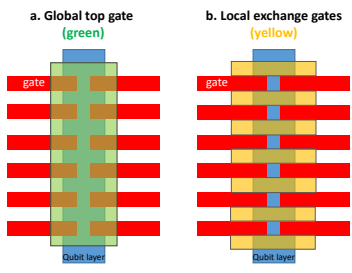
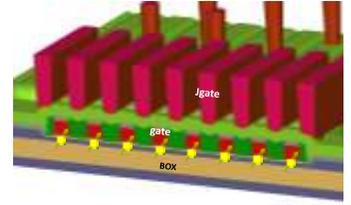

Fig. 1: (left) cross sectional TEM image of a 3-gates Si quantum device on SOI including a global top gate embedded at the M1 level. (right) Drain current (Id) vs top gate voltage (Vtop) electrical characterization. The modulation is very weak for the various front-gate pitch configurations probed.

Fig. 2: Top view layout of a 2xN linear gates array including (a) a global top gate vs (b) local exchange gates intertwined with the front gates.

Fig. 3: 3D schematics of a 2x8 Si qubits array on SOI with embedded metallic trenches as local J-gates.

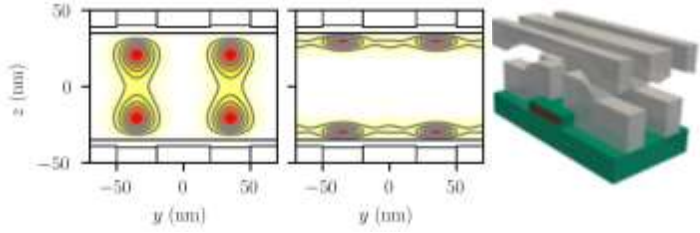
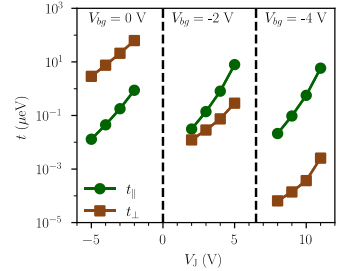
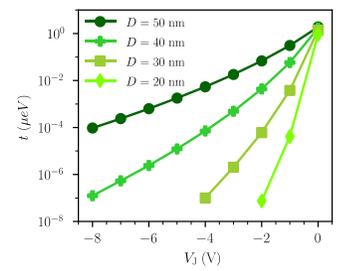

Fig. 4: (right) Simulated periodic 2×2 F2F structure with a 1st level of front gates (gray) partly overlapping the 70 nm wide Si channel (red), and a 2nd level of intertwined J-gates running over the whole width. All gates are 30 nm long. Front gates are $D$ = 50 nm thick. The substrate below the 145 nm thick buried oxide (partly outlined in green) is used as a back gate. (left) Maps of squared electron wave functions in a horizontal cross-section plane, highlighting readout (dots coupled F2F) and two-qubit gates (dots coupled longitudinally) operating points.

Fig. 5: Calculated tunnel couplings $t_{//}$ between neighboring dots and $t_\perp$ between F2F dots as a function of the exchange gate voltage $V_J$ and back gate voltage $V_{bg}$. The front gate voltage is $V_{fg}$ = 50 mV.

Fig. 6: Calculated $t_{/}(V_J)$ for different front gate thicknesses $D$ ($V_{bg}$ = −0.55 V). In these simulations, the exchange gate is separated from the channel by $D$+35 nm of $Si_3N_4$ (no trench etched in the CESL).

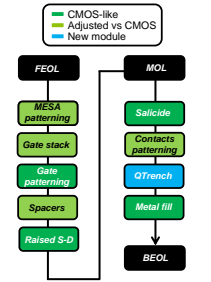
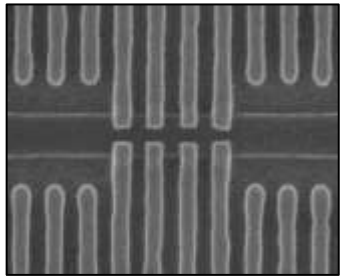
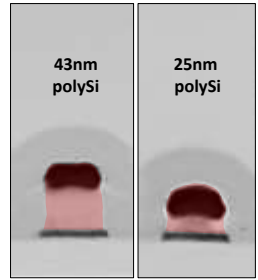
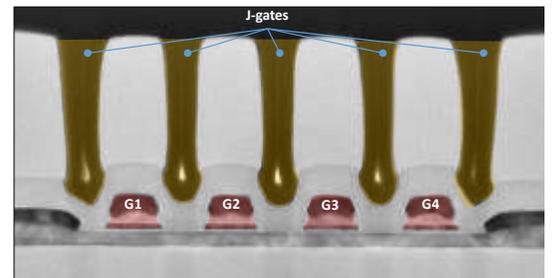

Fig. 7: Integration flow highlighting the minor deviations made compared to standard CMOS fabrication steps.

Fig. 8: SEM top view image of a 2x4 linear gates array at 80nm pitch over a SOI film. Image taken after gate patterning. A full immersion-DUV lithography was used.

Fig. 9: Cross sectional TEM images showing (left) unoptimized and (right) optimized gate stack height yielding a better J-gate efficiency.

Fig. 10: Cross-sectional TEM image of a 80nm pitch 4-gates linear array on SOI with optimized front-gate stack and intertwined 80nm pitch J-gates. Image taken after J-gate patterning. A strict control on overlay and etch selectivity guarantees J-gates functionality.

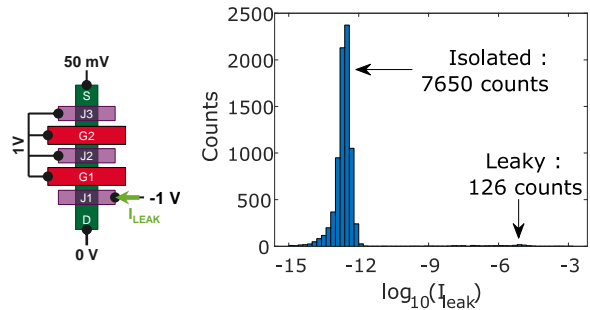
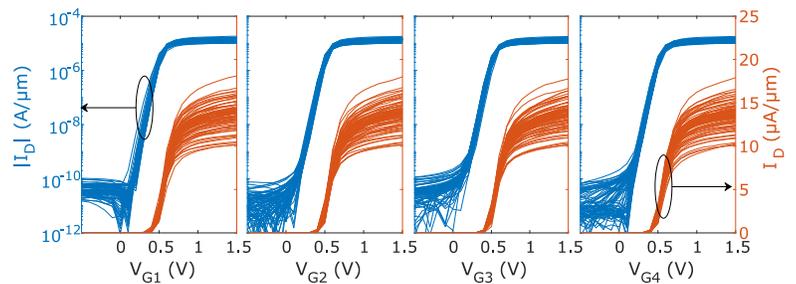

Fig. 11: (left) Schematic of the 2-QDs device bias configuration, for J1 current leakage assessment. Bias and current measurements permutations are done between the 3 J-gates to get the whole dataset. (right) Histogram of the 2-QDs J-gate leakage current distribution. J-gates are considered to be isolated when $I_{LEAK}$<10pA.

Fig. 12: in logaritmic scale (blue curves, left axis) and linear scale (orange curves, right axis) Drain current $I_D$ versus gate voltage $V_G$ curves for 79 functional 4-QDs in series, at $V_{DS}$ = 50 mV, an unswept serie gate bias of 1.5 V and J-gates set to 0V. (functionality criteria : all gates must have $I_{D,max}$ > 10 μA/μm, $I_{D,min}$ < 10 pA/μm, $I_{G,max}$ <10pA/μm and $V_T \in$ [0.3; 0.48] V).

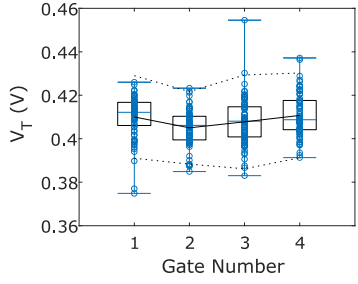
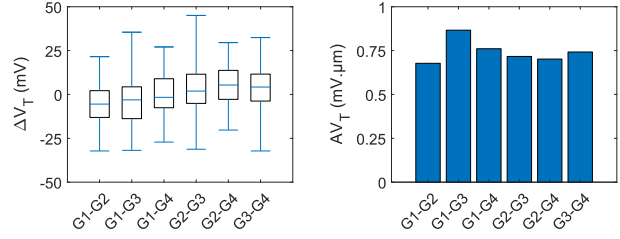
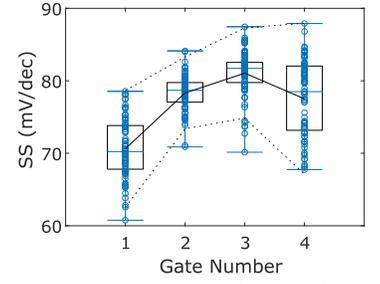

Fig. 13: Threshold voltage distribution boxplot, mean value $\mu_{VT}$ (black line) and 2-$\sigma$ interval (dotted lines) for each of the 4-QDs gates.

Fig. 14: (left) $\Delta V_T$ distribution boxplots and (right) mismatch parameter $AV_T = \sigma(\Delta V_T)\cdot(W\cdot L)^{0.5}$ values (W = 82 nm, L = 45 nm) for all possible gate pairs.

Fig. 15: Subthreshold slope distribution boxplots, mean value $\mu_{VT}$ (black line) and 2-$\sigma$ interval (dotted lines) for each of the 4-QDs gates.

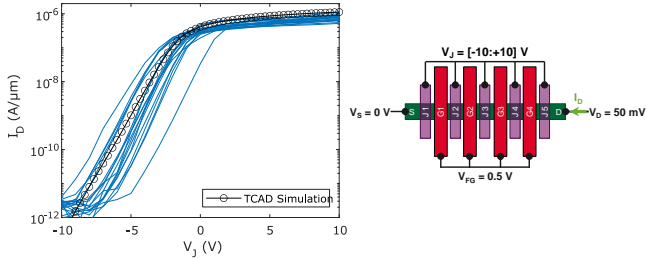
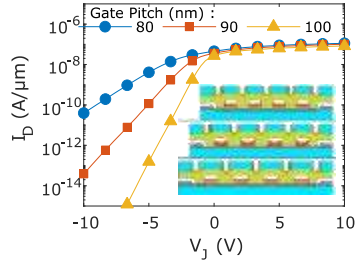
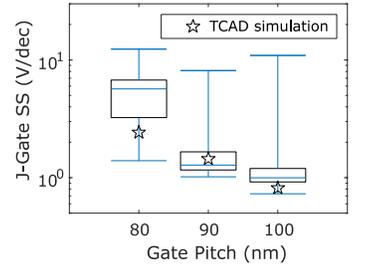

Fig. 16: (left) Measured and simulated drain current $I_D$ versus J-gate voltage $V_J$. (right) Schematic of the bias configuration.

Fig. 17: TCAD simulated drain current versus J-gate voltage for different gate pitches. Inset: simulated device geometries.

Fig. 18: J-gate subthreshold slope distribution boxplots measured on 4-QDs devices and simulated values from Fig. 17, for different gate pitches.

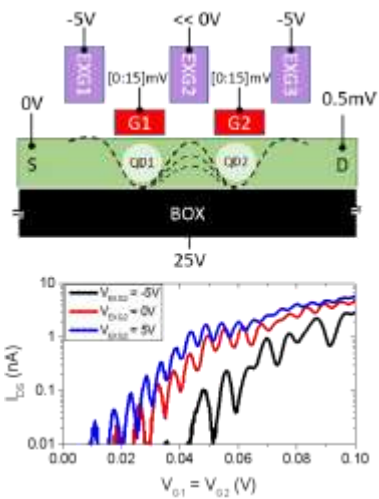
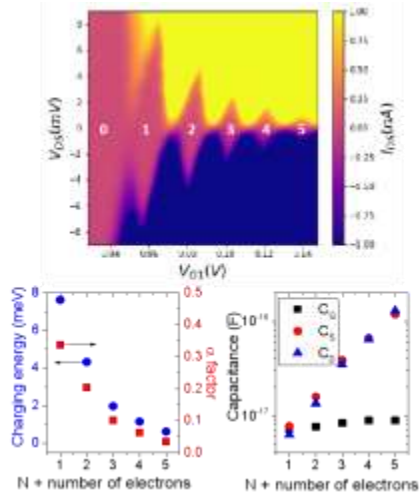
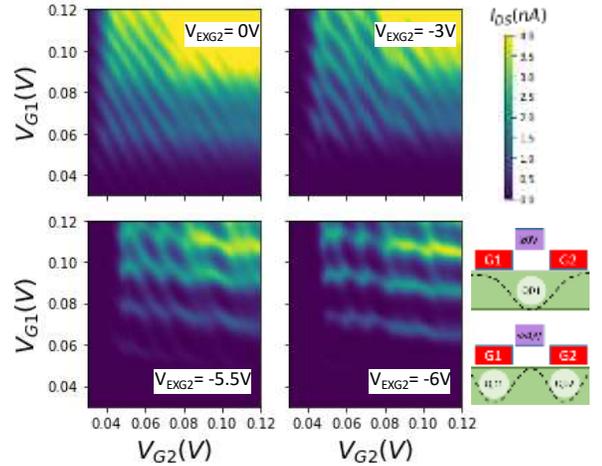

Fig. 19: (top) Schematics of the bias settings used to have the 2-gate devices working as a double QD. J2 enables to control the electrostatic coupling between QD1 and QD2. (bottom) $I_{DS}$-$V_{GS}$ measured at 4.2K. Both G1 and G2 are swept together.

Fig. 20: (top) Coulomb blockade diamonds from QD1, measured at $V_{J2}$ = -7V and $V_{G2}$ = 0.15V. (bottom) Extracted $E_C$, $\alpha$, and gate, source and drain capacitances for each of the 5 first diamonds.

Fig. 21: Stability diagrams measured at the bias conditions indicated in Fig. 19, where $V_{J2}$ varies from 0V (single-dot) down to highly negative values (double-dot regime). When $V_{J2} \geq 0V$, diagonal lines indicate that the QD is similarly coupled to G1 and G2, suggesting it is between both gates (i.e. below J2, as in the schematics on the right). The combination of a back bias and the use of exchange gates yields a high electrostatic control over the QDs, enabling us to operate the devices in different regimes (controlling the QD/reservoir and QD1/QD2 couplings).